\documentclass[12pt]{article}

\input{tcilatex}

\begin{document}

\title{\textbf{The Cosmological Constant Problem and Kaluza-Klein Theory} \ \ }
\date{}
\author{Paul S. Wesson$^{1,2}$ and Hongya Liu$^{3}$}
\maketitle

\bigskip


\bigskip \bigskip $^{1}$Department of Physics, University of Waterloo,
Waterloo, Ontario \ N2L 3G1, Canada. [wesson@astro.uwaterloo.ca].

\bigskip $^{2}$California Institute for Physics and Astrophysics, 360
Cambridge, Ave., Palo Alto, California \ 94306, U.S.A.

$^{3}$Department of Physics, Dalian University of Technology, Dalian 116024,
China. \ [hyliu@dlut.edu.cn]\\

 Correspondence to P. Wesson at address (1).
\newpage

\bigskip \underline{\Large\bf{Abstract}}

We present technical results which extend previous work and show that the
cosmological constant of general relativity is an artefact of the reduction
to 4D of 5D Kaluza-Klein theory (or 10D superstrings and 11D supergravity).
\ We argue that the distinction between matter and vacuum is artificial in
the context of ND field theory. \ The concept of a cosmological ``constant''
(which measures the energy density of the vacuum in 4D) should be replaced
by that of a series of variable fields whose sum is determined by a solution
of ND field equations in a well-defined manner.

\bigskip

\section{\protect\underline{Introduction}}

\qquad The cosmological constant problem in 4D field theory can shortly be
stated as a mismatch in energy densities between those predicted for
particle interactions and that measured by astrophysics and cosmology [1-4].
\ The former energies include those due to zero-point fields and
contributions from the quantum-mechanical vacuum; while the latter is the
energy density of the vacuum in general relativity, and (up to absorbable
constants) is given by the cosmological constant $\Lambda $. \ Following
earlier work [5], it was recently shown that when the 5D Kaluza-Klein field
equations are reduced to the 4D Einstein equations, a cosmological constant
appears which is fixed in size by a length parameter in the metric but has
no more fundamental significance [6,7]. \ It is by now widely known that the
15 Kaluza-Klein equations for apparent vacuum can be rewritten as 10
Einstein equations with matter, plus 4 Maxwell or conservation equations,
and 1 wave equation in a scalar field which augments gravity and
electromagnetism [3,5,6,7]. \ It is in fact always possible in Riemannian
geometry to locally embed an\ ND manifold in an (N+1)D Ricci-flat manifold
[8-10]. \ This implies that the appearance of an artefact like $\Lambda $ is
generic in the reduction to 4D general relativity of 5D Kaluza-Klein theory
[11], 10D superstrings [12] and 11D supergravity [13]. \ In what follows, we
will extend previous work [6-7] to show that the sign of $\Lambda $ depends
on the signature of the 5D metric, and that its size depends on parameters
in the metric. \ Our conclusion will be that when an $N\geq 5$ theory is
reduced to give an $N=4$ energy-momentum tensor, the latter in general
contains matter and vacuum parts whose distinction is artificial. \ It
therefore makes little sense to talk of a cosmological ``constant''. \
Rather, there are contributions to the energy density which are in general
variable, but whose sum is determined by a solution to the field equations
in a well-defined manner.

\section{\protect\underline{The Cosmological ``Constant'' in 5D}}

\qquad In this section, we concentrate on 5D Kaluza-Klein theory, which is
commonly regarded as the low-energy limit of higher-dimensional theories and
can be embedded in them. \ The aim is to see what can be said about the
reduction of the Kaluza-Klein equations

\begin{equation}
R_{AB}=0
\end{equation}
to the Einstein equations

\begin{equation}
R_{\alpha \beta }-\frac{Rg_{\alpha \beta }}{2}-\Lambda g_{\alpha \beta
}=T_{\alpha \beta }\;\;,
\end{equation}
and especially the cosmological constant $\Lambda $. \ Here $R_{AB}$ is the
5D Ricci tensor $(A,B=0,123,4)$; and $R_{\alpha \beta }$, $R$, $g_{\alpha
\beta }$ and $T_{\alpha \beta }$ are the 4D Ricci tensor, Ricci scalar,
metric tensor and energy-momentum tensor $(\alpha ,\beta =0,123)$. \ We use
units which render $c=1$, $8\pi G=1$. \ It is well known that we can move
the last term on the l.h.s. of (2) to the r.h.s., where it determines an
energy density and pressure for the 4D vacuum via

\begin{equation}
\rho =-p=\Lambda \;\;.
\end{equation}
It is also known [3,5] that (1) contain (2), provided we move $\Lambda $ and
define

\begin{eqnarray}
T_{\alpha \beta } &\equiv &\frac{\Phi _{\alpha ;\beta }}{\Phi }-\frac{%
\varepsilon }{2\Phi ^{2}}\QOVERD\{ . {\stackrel{\ast }{\Phi }\;\stackrel{%
\ast \;\;\;}{g_{\alpha \beta }}}{\Phi }-\stackrel{\ast \ast \;}{g_{\alpha
\beta }}+\;g^{\lambda \mu }\;\stackrel{\ast \;\;\;}{g_{\alpha \lambda }}\;%
\stackrel{\ast \;\;}{g_{\beta \mu }}  \nonumber \\
&&-\frac{g^{\mu \nu }\;\stackrel{\ast \;\;\;}{g_{\mu \nu }}\;\stackrel{\ast
\;\;\;}{g_{\alpha \beta }}}{2}+\frac{g_{\alpha \beta }}{4}\left. \left[ 
\stackrel{\ast \;\;\;}{g^{\mu \nu }}\;\stackrel{\ast \;\;\;}{g_{\mu \nu }}%
\;+\left( g^{\mu \nu }\;\stackrel{\ast \;\;}{g_{\mu \nu }}\right) ^{2}\right]
\right\} \,\;\;.
\end{eqnarray}
Here the 5D metric is $dS^{2}=g_{AB}dx^{A}dx^{B}$ and contains the 4D metric 
$ds^{2}=g_{\alpha \beta }dx^{\alpha }dx^{\beta }$. \ We have $g_{\alpha
\beta }=g_{\alpha \beta }(x^{A})$, $g_{4\alpha }=0$ and $g_{44}=\varepsilon
\Phi ^{2}(x^{A})$, where $\varepsilon =\pm 1$. \ We have only used 4 of the
5 available coordinate degrees of freedom to set the electromagnetic
potentials $(g_{4\alpha })$ to zero, so the metric is still general. \ The
effective 4D energy-momentum tensor (4) is also general, and is the basis of
induced-matter theory (for reviews see refs. 3 and 11). \ There is now a
large literature on this, but its essence is clear from (4): the source of
the 4D gravitational field in (2) can be derived from the vacuum equations
(1), provided the latter are not restricted by arbitrary symmetry
conditions. \ [In (4), $\Phi _{\alpha }\equiv \partial \Phi /\partial
x^{\alpha }$, $\Phi _{\alpha ;\beta }$ is the 4D covariant derivative of $%
\Phi _{\alpha }$, and $\stackrel{\ast }{\Phi }\equiv \partial \Phi /\partial
\ell $ with $x^{4}=\ell $ etc., so $T_{\alpha \beta }$ depends on $%
g_{44}=\varepsilon \Phi ^{2}$ and $g_{\alpha \beta }=g_{\alpha \beta }\left(
x^{\alpha },\ell \right) $.] \ Many exact solutions of (1) are known, which
have been applied with (4) to systems ranging from cosmological fluids [14]
to elementary particles [15]. \ It is clear from (4) that 4D matter as
derived from 5D geometry is a sum of contributions that depend on the scalar
field ($\Phi $), the 4D metric (g$_{\alpha \beta }$) and the signature of
the 5D metric ($\varepsilon =\pm 1$). \ Although it is already \ apparent
that the exercise is somewhat artificial, let us proceed to try to isolate
the energy density of the vacuum as measured by $\Lambda $.

\qquad In what are termed canonical coordinates we write $g_{\alpha \beta
}=\left( \ell ^{2}/L^{2}\right) \tilde{g}_{\alpha \beta 
}$%
\linebreak $\left( x^{\alpha },\ell \right) $ and $g_{44}=\varepsilon \Phi
^{2}=-1$ [6]. \ The latter condition uses up the last degree of freedom
allowed by the metric, so the problem is still general. \ But if we now
impose also $\partial \tilde{g}_{\alpha \beta }/\partial \ell =0$, we find
that (2) are satisfied with a cosmological constant

\begin{equation}
\tilde{\Lambda}=\frac{3}{L^{2}}\;\;\;\;\;\;\;.
\end{equation}
Here $L$ is a constant length, introduced to the metric for dimensional
consistency. \ If $L$ is large, then $\tilde{\Lambda}$ is small as required
by cosmology [3,4]. \ However, $\tilde{\Lambda}$ cannot be zero if we
require that the 4D part of the 5D metric be finite in the solar system
[3,7]. \ For then (1) are satisfied with $\partial \tilde{g}_{\alpha \beta
}/\partial \ell =0$ and $\tilde{\Lambda}=3/L^{2}$ by

\begin{eqnarray}
dS^{2} &=&\frac{\tilde{\Lambda}\ell ^{2}}{3}\left\{ \left[ 1-\frac{2M}{r}%
\;-\;\frac{\tilde{\Lambda}r^{2}}{3}\right] dt^{2}-\left[ 1-\frac{2M}{r}-%
\frac{\tilde{\Lambda}r^{2}}{3}\right] ^{-1}dr^{2}-r^{2}d\Omega ^{2}\right\} 
\nonumber \\
&&\;\;\;\;\;\;\;\;\;\;\;\;\;\;\;\;\;\;\;\;\;\;\;\;\;\;\;\;\;\;\;\;\;\;\;\;\;%
\;\;\;\;\;\;\;\;\;\;\;\;\;\;\;\;\;\;\;\;\;\;\;\;\;\;\;-d\ell ^{2}\;\;\;.
\end{eqnarray}
Here $M$ is the mass and $d\Omega ^{2}\equiv \left( d\theta ^{2}+\sin
^{2}\theta d\phi ^{2}\right) $, so this is a 5D embedding for the 4D
Schwarzschild solution. \ It is known that geodesic motion for the 5D metric
(6) reproduces that for the embedded 4D metric [3,6] so there is no way to
differentiate them using the classical tests of relativity [4]. \ We note in
passing that the imaginary transformation $t\rightarrow it$, $r\rightarrow ir
$, $\ell \rightarrow i\ell $, $M\rightarrow iM$ reproduces (6) with the
opposite signs for $\tilde{\Lambda}$ and the last term in the metric. \ We
will return to the signature of the metric below, where we will inquire
whether as in (5) we need to have $\tilde{\Lambda}>0$ or are allowed to have 
$\tilde{\Lambda}<0$.

\qquad In the preceding paragraph, we took the 4D part of the 5D metric as $%
g_{\alpha \beta }=\left( \ell ^{2}/L^{2}\right) \;\tilde{g}_{\alpha \beta
}(x^{\alpha })$ and effectively used $\tilde{g}_{\alpha \beta }$ to define $%
\tilde{\Lambda}=3/L^{2}$ in (5). \ However, if we use $g_{\alpha \beta }$
instead, we find $R_{\alpha \beta }=-3g_{\alpha \beta }/\ell ^{2}$. \ This
describes an empty spacetime with a cosmological constant

\begin{equation}
\Lambda =\frac{3}{\ell ^{2}}\;\;\;\;\;\;\;\;\;.
\end{equation}
Here, $\ell $ is the fifth coordinate, and in the static limit the
correspondence between the energy of a test particle in 4D and 5D requires
the identification $\ell =m$ where $m$ is the rest mass [3,7]. \ Thus (7)
implies that each particle of mass $m$ determines its own $\Lambda $. \ This
is Machian [16,17]; but does not qualify $\Lambda $ of (7) to be called the
cosmological ``constant''. \ The ambiguity between $\tilde{\Lambda}$ of (5)
and $\Lambda $ of (7) is connected technically with whether we use $\tilde{g}%
_{\alpha \beta }$ or $g_{\alpha \beta }$ to raise and lower indices, and
practically with whether a 4D observer experiences the ``pure'' ($\ell $%
-independent) 4D metric or the ``mixed'' ($\ell $-dependent) 4D part of a 5D
metric. \ We defer a consideration of this question, because it will be seen
by what follows to become moot.

\qquad The signature of the 5D metric in Kaluza-Klein theory has important
implications for the sign of the cosmological constant. \ In older work, the
signature was often taken to be (+ - - - -). \ \ However, in modern work it
is frequently left general via (+ - - - $\varepsilon $); and there are
well-behaved solutions with good physical properties which describe waves in
vacuum [18] or galaxies in clusters [19] which have signature (+ - - - +). \
Let us consider a situation similar to those above, but now with a 5D metric 
$dS^{2}=\left( \ell ^{2}/L^{2}\right) \tilde{g}_{\alpha \beta }\left(
x^{\alpha },\ell \right) \;dx^{\alpha }dx^{\beta }+\varepsilon d\ell ^{2}$
which contains a 4D metric $ds^{2}=\tilde{g}_{\alpha \beta }dx^{\alpha
}dx^{\beta }$ which we restrict as before via $\partial \tilde{g}_{\alpha
\beta }/\partial \ell =0$. \ For this, the non-vanishing 5D Christoffel
symbols of the second kind in terms of their counterparts in 4D may be shown
to be

\begin{equation}
^{(5)}\Gamma _{\beta \gamma }^{\alpha }=^{(4)}\Gamma _{\beta \gamma
}^{\alpha },\;\;\;^{(5)}\Gamma _{\beta 4}^{\alpha }=\ell ^{-1\;\;(4)}\delta
_{\beta }^{\alpha },\;\;^{(5)}\Gamma _{\alpha \beta }^{4}=-\varepsilon \ell
L^{-2}\tilde{g}_{\alpha \beta }\;\;\;\;.
\end{equation}
 Using these, we calculate the components of the Ricci tensor as

\begin{equation}
^{(5)}R_{\alpha \beta }=\;\;^{(4)}R_{\alpha \beta }-3\varepsilon L^{-2}%
\tilde{g}_{\alpha \beta },\;\;\;^{(5)}R_{4\alpha
}=0,\;\;\;^{(5)}R_{44}=0\;\;\;.
\end{equation}
These with the field equations (1) give $R_{\alpha \beta }=-\tilde{\Lambda}%
\tilde{g}_{\alpha \beta }$ with a cosmological constant

\begin{equation}
\tilde{\Lambda}=-\frac{3\varepsilon }{L^{2}}\;\;\;\;\;\;.
\end{equation}
We see that if $\varepsilon =-1$ and the fifth dimension is spacelike, then
as above in (5) $\Lambda >0$. \ While if $\varepsilon =+1$ and the fifth
dimension is timelike, then $\Lambda <0$. \ This raises the intriguing
possibility that we could determine the signature of the
(higher-dimensional) world if we could determine the sign of the
cosmological constant.

\qquad Let us now turn our attention from the field equations to the
equations of motion. \ The latter are commonly derived from the variational
principle, which in 5D is writeable symbolically as $\delta \left[ \int dS%
\right] =0$, and leads to the 5D geodesic equation

\begin{equation}
\frac{d^{2}x^{A}}{dS^{2}}\;+\;^{(5)}\Gamma _{BC}^{A}\;\;\frac{dx^{B}}{dS}\;\;%
\frac{dx^{C}}{dS}\;\;=\;\;0\;\;\;\;.
\end{equation}
This has been much studied (see ref. 3 for a bibliography); but here we
follow a new method using as above the metric $dS^{2}=\left( \ell
^{2}/L^{2}\right) ds^{2}+\varepsilon d\ell ^{2}$ with $ds^{2}=\tilde{g}%
_{\alpha \beta }(x^{\gamma })dx^{\alpha }dx^{\beta }$. \ The first of these
relations can be written as

\begin{equation}
\frac{\ell ^{2}}{L^{2}}\left( \frac{ds}{dS}\right) ^{2}+\varepsilon \left( 
\frac{d\ell }{dS}\right) ^{2}=1\;\;\;\;\;.
\end{equation}
Taking $d/dS$ of this we get

\begin{equation}
\frac{\ell ^{2}}{L^{2}}\;\frac{ds}{dS}\;\frac{d^{2}s}{dS^{2}}\;\ +\;\;\frac{%
\ell }{L^{2}}\;\frac{d\ell }{dS}\left( \frac{ds}{dS}\right) ^{2}+\varepsilon 
\frac{d\ell }{dS}\frac{d^{2}\ell }{dS^{2}}\;\;=\;\;0\;\;.
\end{equation}
However, the $A=4$ component of (11) gives with (8) the motion in the extra
dimension as

\[
\frac{d^{2}\ell }{dS^{2}}\;=\;-\;^{(5)}\Gamma _{BC}^{4}\;\frac{dx^{B}}{dS}\;%
\frac{dx^{C}}{dS}\;\;=\;\;-\;^{(5)}\Gamma _{\beta \gamma }^{4}\;\frac{%
dx^{\beta }}{dS}\;\;\frac{dx^{\gamma }}{dS} 
\]

\begin{equation}
\;\;\;\;\;=\;\;\frac{\varepsilon \ell }{L^{2}}\;\tilde{g}_{\alpha \beta }\;%
\frac{dx^{\alpha }}{dS}\;\;\frac{dx^{\beta }}{dS}\;\;\;=\;\;\frac{%
\varepsilon \ell }{L^{2}}\;\left( \frac{ds}{dS}\right)
^{2}\;\;.\,\;\;\;\;\;\;
\end{equation}
Substituting this into (13) gives

\begin{equation}
\frac{d^{2}s}{dS^{2}}\;\;=\;\;\frac{-2}{\ell }\;\frac{d\ell }{ds}\;\;\left( 
\frac{ds}{dS}\right)
^{2}\;\;.\;\;\;\;\;\;\;\;\;\;\;\;\;\;\;\;\;\;\;\;\;\;\;\;\;\;\;\;\;\;\;\;
\end{equation}
This is a convenient relation. \ We note that it is invariant under $%
S\rightarrow iS$, which connects with the possibility that an acausal 5D
manifold may contain a 4D causal one [20]. \ To proceed, we note that we can
manipulate derivatives and use (15) to write

\begin{equation}
\frac{d^{2}x^{A}}{dS^{2}}\;\;=\;\;\left( \frac{ds}{dS}\right) ^{2}\left[ 
\frac{d^{2}x^{A}}{ds^{2}}\;\;-\;\;\frac{2}{\ell }\;\frac{d\ell }{ds}\;\;%
\frac{dx^{A}}{ds}\right] \;\;\;.
\end{equation}
We can use this with (14) to rewrite the fourth component of the geodesic
equation (11) as

\begin{equation}
\frac{d^{2}x^{4}}{dS^{2}}\;+\;^{(5)}\Gamma _{BC}^{4}\,\;\frac{dx^{B}}{dS}\;%
\frac{dx^{C}}{dS}\;=\;\left( \frac{ds}{dS}\right) ^{2}\;\left[ \frac{%
d^{2}\ell }{dS^{2}}\;-\;\frac{2}{\ell }\;\left( \frac{d\ell }{dS}\right)
^{2}\;-\;\frac{\varepsilon \ell }{L^{2}}\right] \;\;\;.
\end{equation}
The motion is geodesic if

\begin{equation}
\frac{d^{2}\ell }{ds^{2}}\;-\;\frac{2}{\ell }\;\left( \frac{d\ell }{ds}%
\right) ^{2}\;-\;\frac{\varepsilon \ell }{L^{2}}\;=\;0\;\;\;\;\;\;\;\;\;\;\;%
\;\;\;\;\;\;\;\;\;\;\;\;\;\;\;\;\;\;\;\;\;\;\;\;\;\;\;\;\;\;\;\;.
\end{equation}
This can be rewritten as

\begin{equation}
\frac{d^{2}}{ds^{2}}\;\left( \frac{1}{\ell }\right) \;+\;\frac{\varepsilon }{%
L^{2}}\;\left( \frac{1}{\ell }\right) \;=\;0\;\;\;\;\;.
\end{equation}
Solutions of this will depend on two arbitrary constants, which we take to
be special values $\ell _{\ast }$ and s$_{\ast }$ of the fifth coordinate
and the 4D interval. \ Then for the choices of $\varepsilon $ we can write
the solutions of (19) as

\[
\ell =\frac{\ell _{\ast }}{\cosh \;\left[ \left( s-s_{\ast }\right) /L\right]
}\;\;\;\;\;\;\;\;\;\;\;\;\;\;\;,\;\;\;\varepsilon =-1 
\]

\begin{equation}
\ell =\frac{\ell _{\ast }}{\cos \;\left[ \left( s-s_{\ast }\right) /L\right] 
}\;\;\;\;\;\;\;\;\;\;\;\;\;\;\;\;\;,\;\;\;\;\varepsilon =+1\;.
\end{equation}
We see that the motion in the fifth dimension depends on the signature. \
And if $L$ is related to $\Lambda $ via a relation like (5) or (6), it also
depends on the cosmological constant. \ Further, if $\ell $ is related to
the rest mass $m$ of a particle [3, 7, 21] then the latter may either
increase or decrease with cosmological time depending on the signs of $%
\varepsilon $ and $\Lambda $. \ These possibilities naturally lead us to
wonder if the motion in 4D spacetime is geodesic or not. \ To answer this,
we use (8) and (16) in the expanded version of (11), which then gives

\begin{equation}
\frac{d^{2}x^{\alpha }}{dS^{2}}\;\;+\;\;^{(5)}\Gamma _{BC}^{\alpha }\,\;%
\frac{dx^{B}}{dS}\;\frac{dx^{C}}{dS}=\left( \frac{ds}{dS}\right) ^{2}\left[ 
\frac{d^{2}x^{\alpha }}{ds^{2}}+^{\left( 4\right) }\Gamma _{\beta \gamma
}^{\alpha }\frac{dx^{\beta }}{ds}\;\frac{dx^{\gamma }}{ds}\right] .
\end{equation}
Clearly, if the 5D motion is geodesic then so is the 4D motion. \ This
result agrees with others in the literature [3, 6, 7, 20, 21] and is
remarkable: questions to do with $\varepsilon $, $\Lambda $ and $m$ are
confined to the fifth dimension and the motion is the standard kind in the
four dimensions of spacetime.

To here, we have concentrated on elucidating the nature of the cosmological
constant by examining the field equations and the equations of motion for
metrics of canonical form. \ We have seen that there are, from the 5D
perspective, several different ways to define this parameter. \ From the
field equations, and particularly $R_{\alpha \beta }$, we can obtain
relations like (5) and (7), which modulo a conformal factor in the 4D metric
are equivalent mathematically. \ However, they are different physically. \
>From the equations of motion, we can obtain relations like (20) which
implicate the cosmological constant in the fifth component of the geodesic
but leave the four spacetime components (21) the same as in general
relativity, which means by (6) that there is a cosmological force $%
\widetilde{\Lambda }\;r/3$ that acts in the solar system and other 1-body
systems. \ However, unlike in general relativity, the 4D part of the 5D
metric is only finite if $\widetilde{\Lambda }$ is finite. \ In addition to
geometrical and dynamical ways to define the cosmological constant must be
added that which embodies the equation of state, which in Einstein theory is
(3). \ This brings us back to (4), the general expression for the induced
energy-momentum tensor which is obtained when Einstein's equations (2) are
embedded in the Kaluza-Klein equations (1). \ In the above, we have focused
for algebraic reasons on how to derive the cosmological constant for metrics
in canonical coordinates (6). \ However, modern Kaluza-Klein theory is fully
covariant [3] so the question arises of whether it is possible to identify a
contribution to the $T_{\alpha \beta }$ of (4) that can be uniquely
attributed to a cosmological constant. \ We believe that the answer to this
question is No. \ Some comments to support this are in order. (a) While the
last term in (4) is proportional to $g_{\alpha \beta }$, the coefficient is
essentially the Ricci scalar [5], and so cannot be identified with a
cosmological constant. \ (b) There is no way to tell which if any terms in
(4) may become proportional to $g_{\alpha \beta }$ after substitution of an
exact solution of the field equations. (c) Most physical interpretations of
(4) are made by comparison with a single perfect fluid, but multi-fluid
models (e.g. matter and radiation) with possibly imperfect fluids (e.g. with
viscosity) are more realistic, and it is difficult to see how to extract a
component due to a cosmological constant from there. \ (d) The $T_{\alpha
\beta }$ of (4) satisfies Einstein's equations (2) in a formal manner, but
it contains terms that depend on a scalar field which lies outside general
relativity, so the ``matter'' in (4) can be different to what is
conventional, and the equation of state of the ``vacuum'' can also be
different. \ (e) Indeed, it is clear from an inspection of (4) that there is
no unique way to separate what are conventionally called ``matter'' and
``vacuum'', and while there is currently some discussion about whether the
effects normally attributable to dark matter may be due to a cosmological
constant with the fluid characteristics of (3), the 5D view as formalized by
(4) is that ``matter'' and ``vacuum'' contributions to the energy density
are mixed, the distinction having more to do with history than physics. \
(f) There are exact solutions of the field equations (1) known which with
(4) can be interpreted as involving ordinary matter and a cosmological
``constant'' which is time-dependent [22], but while \ these may help
resolve well-known problems with the matching of cosmological data [23],
they indicate that a cosmological ``constant'' if it is defineable at all
can be a function of the coordinates.

\section{\protect\underline{Conclusion}}

\qquad We have presented a series of technical results involving the
embedding of 4D Einstein theory in 5D Kaluza-Klein theory, which by
extension can be applied to 10D superstrings and 11D supergravity. \ In the
5D case, it is arguable that there is no logical distinction between
``matter'' and ``vacuum'' contributions to the energy density; and since the
vacuum in general relativity can be related to the cosmological constant, we
are lead to seriously doubt if the latter parameter has any real meaning. \
What \underline{does} have meaning is an exact solution of the ND field
equations, which when reduced to 4D defines an induced energy-momentum
tensor whose various and variable terms determine the energy density in a
well-defined manner. \ This implies that the mismatch in energy densities
derived from quantum field theory and general relativity is merely a
consequence of restricting the physics to 4D. \ Put another way, we believe
that the cosmological constant problem does not exist in $N\geq 5D.$

\ \ \ 

\ \ 

\bigskip \underline{\Large\bf{Acknowledgement}}
\qquad We thank B. Mashhoon, J.M. Overduin and W.N. Sajko for comments; and
the Natural Sciences and Engineering Research Council of Canada and the
National Natural Science Foundation of China for support.

\bigskip\ \ \ 


\underline{\Large\bf{References}}
\begin{enumerate}
\item  S. Weinberg, Rev. Mod. Phys. \underline{61}, 1 (1989).

\item  Y.J. Ng, Int. J. Mod. Phys. D\underline{1}, 145 (1992).

\item  P.S. Wesson, Space, Time, Matter, World Scientific Singapore\ (1999).

\item  C.M. Will, Theory and Experiment in Gravitational Physics, Cambridge
Un. Press, Cambridge (1993).

\item  P.S. Wesson, J. Ponce de Leon, J. Math. Phys. \underline{33}, 3883
(1992).

\item  B. Mashhoon, P.S. Wesson, H. Liu, Gen. Rel. Grav. \underline{30}, 555
(1998).

\item  H. Liu, B. Mashhoon, Phys. Lett. A \underline{272}, 26 (2000).

\item  S. Rippl, C. Romero, R. Tavakol, Class. Quant. Grav. \underline{12},
2411 (1995).

\item  C. Romero, R. Tavakol, R. Zalaletdinov, Gen. Rel. Grav. \underline{28}%
, 365 (1996).

\item  J.E. Lidsey, C. Romero, R. Tavakol. S. Rippl, Class. Quant. Grav. 
\underline{14}, 865 (1997).

\item  J.M. Overduin, P.S. Wesson, Phys. Rep. \underline{283}, 303 (1997).

\item  M.B. Green, J.H. Schwarz, E. Witten, Superstring Theory, Cambridge
Un. Press, Cambridge (1987).

\item  P. West, Introduction to Supersymmetry and Supergravity, World
Scientific, Singapore (1986).

\item  P.S. Wesson, Astrophys. J. \underline{394}, 19 (1992).

\item  P.S. Wesson, H. Liu, Phys. Lett B \underline{432}, 266 (1998).

\item  J. B. Barbour, H. Pfister (eds.), Mach's Principle: From Newton's
Bucket to Quantum Gravity, Birkhauser, Boston (1995).

\item  M. Jammer, Concepts of Mass in Contemporary Physics and Philosophy,
Princeton Un. Press, Princeton (2000).

\item  A. Billyard, P.S. Wesson, Gen. Rel Grav. \underline{28}, 129 (1996).

\item  A. Billyard, P.S.Wesson, Phys. Rev. D \underline{53}, 731 (1996).

\item  A. Davidson, D.A. Owen, Phys. Lett. B. \underline{177}, 77 (1986).

\item  H. Liu, B. Mashhoon, Ann. Phys. (Leipzig) \underline{4}, 565 (1995).

\item  W.N.Sajko, Ph.D. Thesis, University of Waterloo (2000).

\item  J.M. Overduin, Astrophys. J. \underline{517}, L1 (1999).
\end{enumerate}

\end{document}